\documentclass{desyproc}

\newcommand\POWHEG{{\tt POWHEG}}
\newcommand\POWHEGBOX{{\tt POWHEG-BOX}}
\newcommand\HERWIG{{\tt HERWIG}}
\newcommand\HWpp{{\tt HERWIG++}}
\newcommand\PYTHIA{{\tt PYTHIA}}
\newcommand\MCatNLO{{\tt MC@NLO}}
\newcommand\matB{{\cal B}}
\newcommand\matR{{\cal R}}
\newcommand\matVb{{\cal V}_{\rm b}}
\newcommand\matSV{{\cal V}}
\newcommand\matMCs{{\cal M}}

\newcommand\muR{\mu_{\sss\rm R}}
\newcommand\NC{N_{\rm c}}
\newcommand\sss{\mathchoice%
{\displaystyle}%
{\scriptstyle}%
{\scriptscriptstyle}%
{\scriptscriptstyle}%
}

\begin{document}
\title{NLO and Parton Showers: the \POWHEGBOX{}}


%
\author{{\slshape Simone Alioli}\\[1ex]
Deutsches Elektronen-Synchrotron DESY, Platanenallee 6, 15738 Zeuthen, Germany }

%

%

\contribID{xy}  
\confID{1964}
\desyproc{DESY-PROC-2010-01}
\acronym{PLHC2010}
\doi            

\maketitle

\begin{abstract}
  We describe the \POWHEGBOX{} package, a general computer code
framework for implementing NLO calculations in Shower Monte Carlo
programs according to the \POWHEG{} method.

\end{abstract}

\section{The \POWHEG{} method}
\label{seq:master}
\emph{Next-to-leading order} (NLO) perturbative QCD computations as
well as \emph{Shower Monte Carlo} (SMC) programs are fundamental tools
for the present-days particle physics phenomenology. In particular,
SMC programs incorporate the description of a generic high-energy
hadronic collision process, starting from the collision between
constituents and developing the parton shower, that increases
the number of final-state particles by means of strongly ordered
subsequent emissions.  Eventually, the interface with a
phenomenological hadronization model, enables the comparison with
experimental data. For these reasons, they are routinely used by
experimentalists to simulate signal and backgrounds processes in
physics searches. Nevertheless, the demand for better and better
predictions from high energy experiments calls for improving the
precision of existing SMC's, including NLO corrections.  The
\MCatNLO{}~\cite{Frixione:2002ik} method has shown first how to reach
NLO accuracy for inclusive quantities, implementing the hard
subprocess at NLO and developing showers within the \emph{leading
  logarithmic approximation}, avoiding double counting of radiation.
In this way one achieves benefits of both approaches: exclusive final
states generation of SMC's and accuracy of NLO calculations.

The \POWHEG{} method is a different  prescription for interfacing NLO
calculations with parton shower generators. It was first suggested in
ref.~\cite{Nason:2004rx}, and was described in great detail in
ref.~\cite{Frixione:2007vw}.  This method is independent from the
Monte Carlo program used for subsequent showering and generates
positive weighted events only.  In these respects it improves the
\MCatNLO{} approach.
 Until now,
the \POWHEG{} method has been successfully applied to several processes, both at lepton~\cite{LatundeDada:2006gx,LatundeDada:2008bv}  and hadron colliders~\cite{Nason:2006hfa,Frixione:2007nw,Alioli:2008gx,Hamilton:2008pd,Papaefstathiou:2009sr,Alioli:2008tz,Hamilton:2009za,Alioli:2009je,Nason:2009ai}.
In these implementations, it has
been interfaced to the \HERWIG{}~\cite{Corcella:2000bw, Corcella:2002jc},
\PYTHIA{}~\cite{Sjostrand:2006za} and \HWpp{}~\cite{Bahr:2008pv} SMC programs.

In the \POWHEG{} method  the hardest radiation\footnote{By hardest we mean the radiation with the highest transverse momentum, either with respect to the beam for initial state radiation (ISR), either with respect to another parton for final state radiation (FSR). } is generated first, independently from the following ones.
Schematically\footnote{Here we avoid entering into the details concerning the radiation regions and the correct treatment of the associated flavour configurations. The interested reader can find further explanations  in the original papers~\cite{Nason:2004rx,Frixione:2007vw}}, the hardest radiation is distributed according to 
\begin{equation}
\label{eq:master}
d\sigma=\bar{B}\left(\Phi_{B}\right)\,d\Phi_{B}\,\left[\Delta_{R}\left(p_{T}^{\min}\right)+
\frac{R\left(\Phi_{R}\right)}{B\left(\Phi_{B}\right)}\,\Delta_{R}\left(k_{T}\left(\Phi_{R}\right)\right)\,d\Phi_{\mathrm{rad}}\right]\,,
\end{equation}
where $B\left(\Phi_{B}\right)$ is the Born contribution and
\begin{equation}
\label{eq:bbar}
\bar{B}\left(\Phi_{B}\right)=B\left(\Phi_{B}\right)+
\left[V\left(\Phi_{B}\right)+\int d\Phi_{\mathrm{rad}}\, R\left(\Phi_{R}\right)\right]
\end{equation}
is the NLO differential cross section at fixed underlying Born kinematics and integrated over the radiation variables. 
The transverse momentum of the emitted parton, with respect to the beam or to another particle, depending on the region of singularity, is  denoted by
 $k_T\left(\Phi_{R}\right)$. 
The lower cutoff $p_{T}^{\min}$ is necessary in order to avoid the coupling constant to reach unphysical values.
$V\left(\Phi_{B}\right)$  and $R\left(\Phi_{R}\right)$ are the virtual and the real corrections  and
in the expression
within the square bracket in Eq.~(\ref{eq:bbar})
a procedure that takes care of the cancellation of
soft and collinear singularities is understood, \emph{e.g.} Frixione-Kunszt-Signer (FKS)~\cite{Frixione:1995ms} or Catani-Seymour (CS) dipole subtraction.
Then, 
\begin{equation}
\Delta_{R}\left(p_{T}\right)=\exp\left[-\int d\Phi_{\mathrm{rad}}\,\frac{R\left(\Phi_{R}\right)}{B\left(\Phi_{B}\right)}\,\theta\left(k_{T}\left(\Phi_{R}\right)-p_{T}\right)\right]\,
\end{equation}
is the \POWHEG{} Sudakov, that is the probability of not having an emission harder that $p_{T}$.
Equation~(\ref{eq:master}) can be seen as an improvement on the original SMC hardest-emission formula, since the Born cross section is replaced with $\bar{B} \left(\Phi_{B}\right)$ which is  normalized to NLO by construction. 
At small transverse momenta the \POWHEG{} Sudakov becomes equal to a standard SMC one. However, the NLO accuracy of Eq.~(\ref{eq:master}) is  maintained for inclusive quantities. 
Moreover, the high$-p_T$ radiation region is correctly described by the real contributions 
\begin{equation}
d\sigma \approx \bar{B}\left(\Phi_{B}\right)\,d\Phi_{B}\,
\frac{R\left(\Phi_{R}\right)}{B\left(\Phi_{B}\right)} d\Phi_{\mathrm{rad}}\  \approx R \left(\Phi_{R}\right) d\Phi_{B} d\Phi_{\mathrm{rad}} \, ,
\end{equation}
since  $\Delta_{R} \approx 1$ and $ \bar{B} /B \approx  1 + \mathcal{O} \left({\alpha_{\scriptstyle{\rm s}}
 } \right) $. 
After having generated the hardest radiation, one can interface with any available  shower generator, in order to develop the rest of the shower.
To avoid the double-counting, the SMC is required to be either $p_T-$ordered or to have the ability to veto emissions with a $p_T$ harder than the first one\footnote{
All modern SMC generators compliant with the \emph{Les Houches Interface for User Processes}~\cite{Boos:2001cv}  implement this feature.}.

\section{The \POWHEGBOX{}}

In a  real collision process several colored massless partons are present, either in the initial or the final state. One thus should repeat the procedure outlined in Sec. \ref{seq:master} for every possible singular region, associated with any  massless colored leg becoming collinear to another one, or soft.
In order to do this, the full real emission cross section is decomposed into a sum of terms, each of which has at most one collinear and one soft singularities. The radiation is then generated independently in each of this regions,
but only the hardest radiation is retained and the event is generated according to the flavour and kinematics associated to it.
Because of this complexity, an automatic tool, the \POWHEGBOX{}, has been built~\cite{Alioli:2010xd}, in order to help the inclusion of new processes. 
On the other hand, the
 \POWHEGBOX{} may also be seen as a library, where previously implemented processes are available in a common framework.

The user wishing to include a new NLO calculation must
only know how to communicate the needed information to the 
\POWHEGBOX{}. This happens either defining the appropriate variables, either providing  the necessary \texttt{Fortran} routines. The required inputs are: 
\begin{enumerate}
\item The number of legs in Born process, \emph{e.g.} \texttt{nlegborn}$=5$ for $pp\to(Z\to e^+ e^- ) + j$.
\label{enum1}
\item The list of Born and Real processes flavours
\texttt{flst\_born(k=1..nlegborn,j=1..flst\_nborn)},
 \texttt{flst\_real(k=1..nlegreal,j=1..flst\_nreal)}, 
 according to PDG~\cite{PDG} conventions\footnote{Internally gluons  are labelled $0$ instead of the PDG value of $21$. At the moment of writing the event on the Les Houches common block, the PDG value is restored.}. Flavor is defined incoming (outgoing) for incoming (outgoing) fermion lines, \emph{e.g.} \texttt{[5,2,23,6,3,0]} for $ b u\ \to\ Z t s g$.
\label{enum2}
\item The Born phase space{ \texttt{Born\_phsp(xborn)}}, given \texttt{xborn(1...ndims)} random numbers in the unit \texttt{ndims}-dimensional hypercube. This routine should also set the Born phase space Jacobian {\texttt{kn\_jacborn}}, the Born momenta \texttt{kn\_pborn}, \texttt{kn\_cmpborn} in lab. and CM frames and the Bjorken $x$'s, \emph{i.e.} \texttt{kn\_xb1,kn\_xb2} .
\label{enum3} 
\item The routines that performs the initialization of the couplings, \texttt{init\_couplings}, and  the setting of the factorization and renormalization scales, \texttt{set\_fac\_ren\_scales(muf,mur)}.
\label{enum4}
\item The Born squared amplitudes $ \matB = |\matMCs|^2$, summed and averaged over color and helicities as well as the color-ordered Born squared amplitudes $ \matB_{jk}$ and the helicity correlated Born squared amplitudes $ \matB_{k,\mu\nu}$, where $k$ runs over all external gluons. 
The corresponding user routine is defined by  
 \texttt{setborn(p,bflav,born,bornjk,bornmunu)},
for momenta \texttt{p(0:3,1:nlegborn)} and flavour string \texttt{bflav(1:nlegborn)} .
\label{enum5}
\item The real emission squared amplitudes $ \matR $ user routine 
  \texttt{setreal(p,rflav,amp2)}, 
for momenta {\texttt{p(0:3,1:nlegreal)}} and flavour string { \texttt{rflav(1:nlegreal)}} .
\label{enum6}
\item The finite part of the interference of Born and virtual amplitude contributions $\matVb = 2 \mathrm{Re} \{ \matB \times \matSV \} $, after factorizing out 
  the common factor $ \mathcal{N} = \frac{(4 \pi)^{\epsilon}}{\Gamma \left(1 - \epsilon\right)} \left(\frac{\muR^2}{Q^2} \right)^{\epsilon}$.  The user routine is  called 
 \texttt{setvirtual(p(0:3,1:nlegborn),vflav(1:legborn),virtual)}.
\label{enum7}
\item The Born color structures in the large $\NC$ limit,   set by the routine \texttt{borncolour\_lh}   and passed through the Les Houches interface~\cite{Boos:2001cv}. 
\label{enum8}
\end{enumerate} 
We remark that items (\ref{enum1}-\ref{enum7}) are the
 usual ingredients needed to perform a NLO calculation in any subtraction method.  Item (\ref{enum8}) is instead needed to provide a defined color structure to the SMC generator.
Internally, the \POWHEGBOX{} implements the FKS subtraction procedure in a general way. At the beginning, it automatically evaluates the combinatorics, identifying all the singular regions and the corresponding underlying Born contributions. It also performs the projection of real contributions over  the singular region and computes the subtraction counterterms,  from soft and collinear  approximations of real emissions. Then, it  builds the  ISR and FSR  phase space, according to the FKS parametrization of the singular region and performs the integration.  Eventually, one gets the NLO differential cross section. At this stage, one can also interface to some analysis routine to obtain NLO differential distributions as a byproduct. After the integration stage, it performs the calculation of upper bounds for an efficient generation of Sudakov-suppressed events and then the generation of hardest radiation, according to the \POWHEG{} Sudakov. At this point, the generated event, which contains at most only one extra radiation, has to be passed to a standard SMC program, for developing the rest of the shower. This can be done either on-the-fly or  storing the events on a Les Houches events file~\cite{Alwall:2006yp}.  Standard analysis routines, at partonic and hadronic level, are provided for included processes, as well as drivers for common SMC generators. Users can modify them or implement new ones.

\subsection{Recent developments}
\begin{figure}[hb]
\begin{minipage}[b]{0.45\linewidth}
\centering
\includegraphics[width=0.9\textwidth]{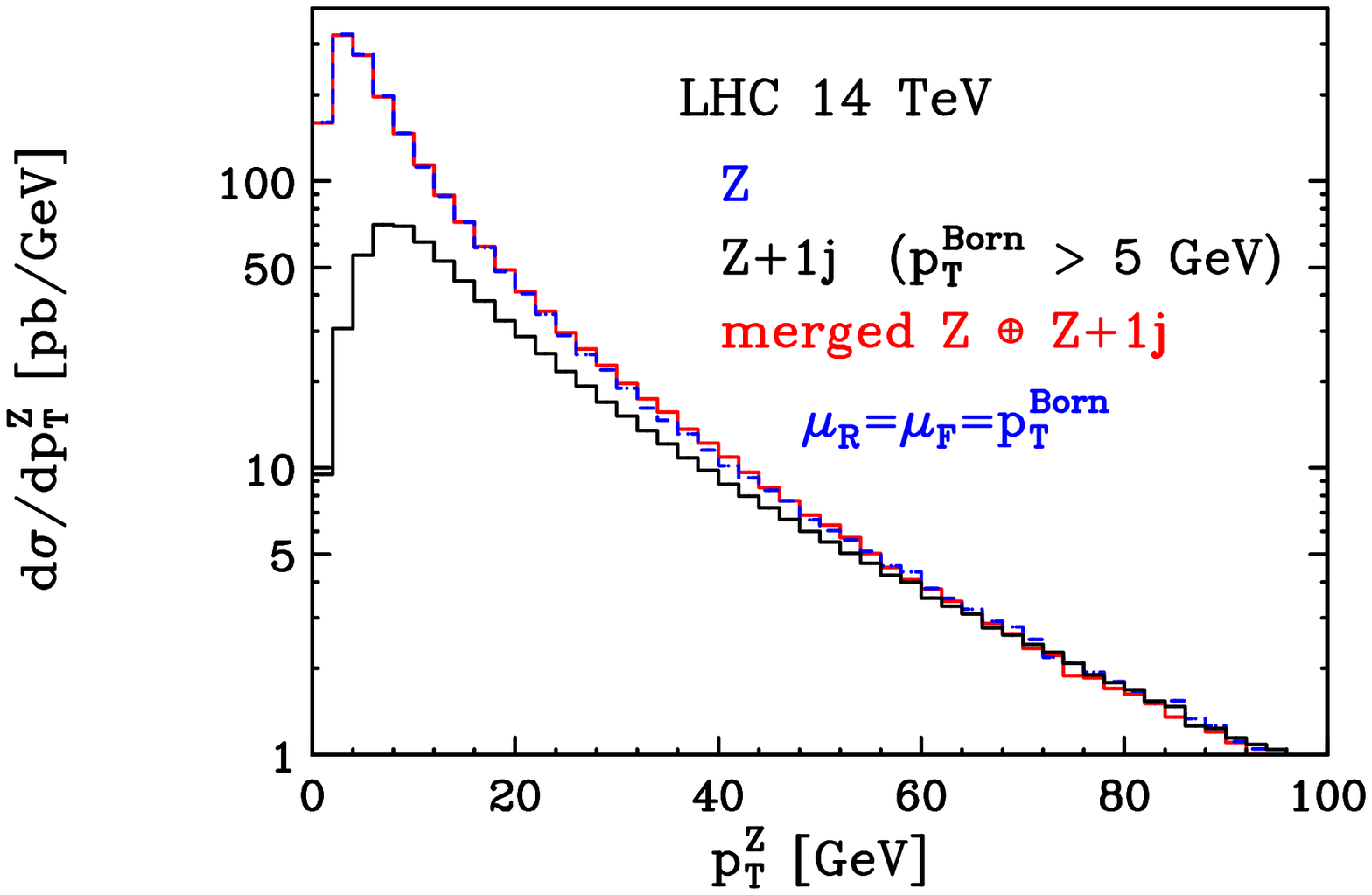}
\caption{The $p_T$ distribution of the $Z$ boson in single $Z$ production (blue dashed curve), in $Z+1$ jet (black solid) and in the merged sample (red solid).}
\label{Fig:1}
\end{minipage}
\hspace{20pt}
\begin{minipage}[b]{0.45\linewidth}
\centering
\includegraphics[width=0.9\textwidth]{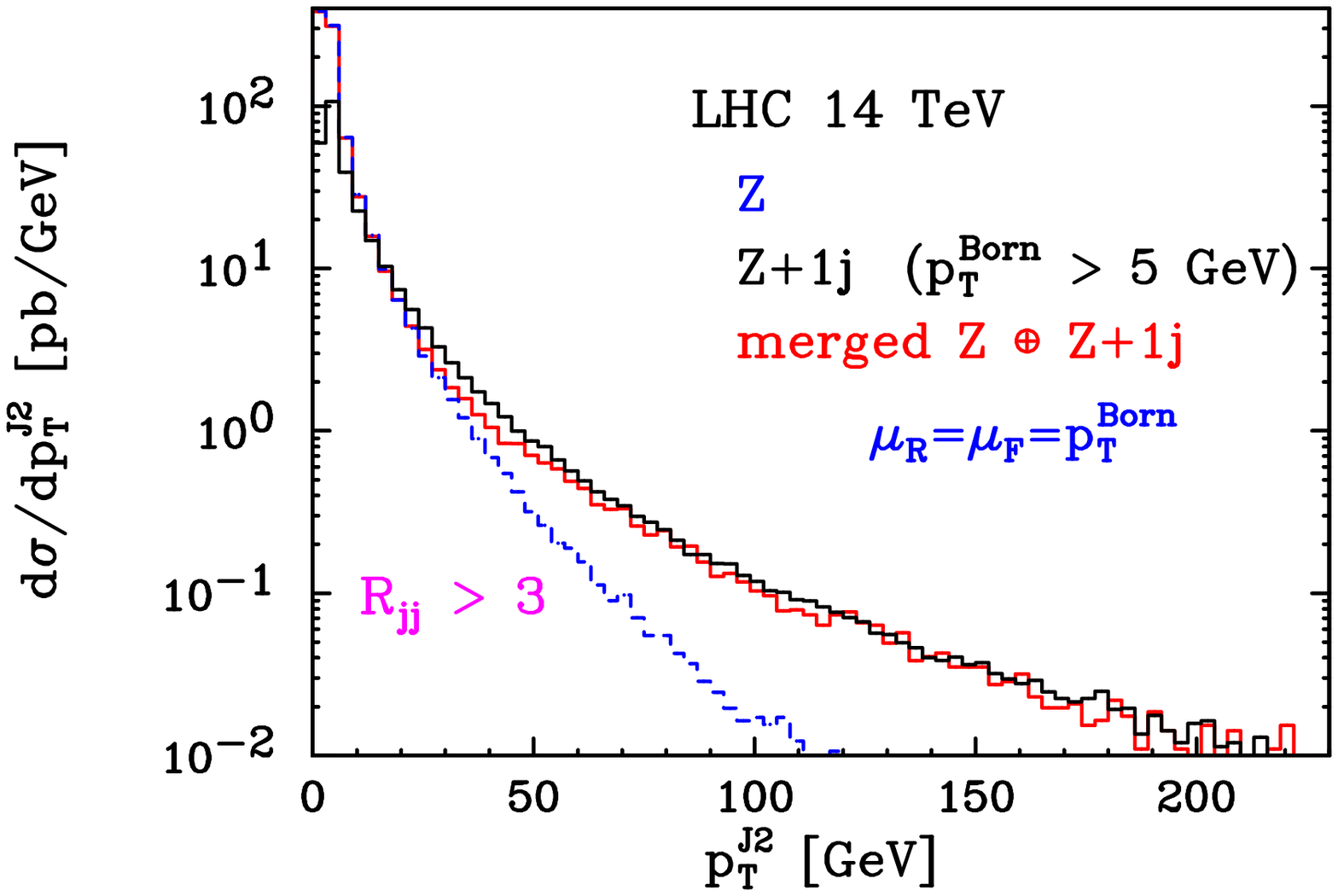}
\caption{The $p_T$ distribution of the \emph{next-to-hardest} jet in single $Z$ production (blue dashed curve), in $Z+1$ jet (black solid) and in the merged sample (red solid).}
\label{Fig:2}
\end{minipage}
\end{figure}

Recently, thanks to this framework, the relatively complex process of $Z+1\ jet$ production has been implemented~\cite{POWHEG_Zjet}.
This is a promising processes for jet calibration with the early LHC data.  It is also an important source of missing energy signal as well as a background to many new physics searches. In experimental studies carried up until now, the NLO theoretical calculations were supplemented by correction factors for shower, hadronization and underlying event effects. However, these factors were evaluated by means of standard LO SMC programs. It is clear the advantage to have a SMC program which is NLO accurate, in order to ease and  improve the comparisons with experimental results.  
We have tried a simple approach~\cite{POWHEG_Zjet} to merge consistently $Z$ and $Z + 1\ jet$ samples, in order to obtain a description as 
smooth and accurate as possible both in the low and high-transverse momentum regions. The results are showed in Figs.~\ref{Fig:1} and~\ref{Fig:2}.

From the two figures, one can see how the merged sample 
models both the single $Z$ Sudakov form factor, that plays an important role in resumming collinear/soft logarithms in the low-$p_T$ region and the high-$p_T$ behaviour of the next-to-hardest jet, which follows the $Z+1$ jet distribution.
In this last figure, jest are reconstructed according to the $k_T$ algorithm, imposing also an angular separation $R_{jj}>3$  .

\section*{Acknowledgments}
All results presented in this talk have been obtained in collaboration with P. Nason, C. Oleari and
E. Re.
This work has been supported in part by the Deutsche
Forschungsgemeinschaft in SFB/TR 9.


\begin{footnotesize}

\end{footnotesize}


\begin{thebibliography}{99}

\bibitem{Frixione:2002ik}
  S.~Frixione and B.~R.~Webber,
  JHEP {\bf 0206} (2002) 029
  [arXiv:hep-ph/0204244].

\bibitem{Nason:2004rx}
  P.~Nason,
  JHEP {\bf 0411} (2004) 040
  [arXiv:hep-ph/0409146].



\bibitem{Frixione:2007vw}
  S.~Frixione, P.~Nason and C.~Oleari,
  JHEP {\bf 0711} (2007) 070
  [arXiv:0709.2092 [hep-ph]].

\bibitem{LatundeDada:2006gx}
  O.~Latunde-Dada, S.~Gieseke and B.~Webber,
  JHEP {\bf 0702} (2007) 051
  [arXiv:hep-ph/0612281].

\bibitem{LatundeDada:2008bv}
  O.~Latunde-Dada,
  Eur.\ Phys.\ J.\  C {\bf 58} (2008) 543
  [arXiv:0806.4560 [hep-ph]].


\bibitem{Nason:2006hfa}
  P.~Nason and G.~Ridolfi,
  JHEP {\bf 0608} (2006) 077
  [arXiv:hep-ph/0606275].

\bibitem{Frixione:2007nw}
  S.~Frixione, P.~Nason and G.~Ridolfi,
  JHEP {\bf 0709} (2007) 126
  [arXiv:0707.3088 [hep-ph]].


\bibitem{Alioli:2008gx}
  S.~Alioli, P.~Nason, C.~Oleari and E.~Re,
  JHEP {\bf 0807} (2008) 060
  [arXiv:0805.4802 [hep-ph]].

\bibitem{Hamilton:2008pd}
  K.~Hamilton, P.~Richardson and J.~Tully,
  JHEP {\bf 0810} (2008) 015
  [arXiv:0806.0290 [hep-ph]].


\bibitem{Papaefstathiou:2009sr}
  A.~Papaefstathiou and O.~Latunde-Dada,
  JHEP {\bf 0907} (2009) 044
  [arXiv:0901.3685 [hep-ph]].

\bibitem{Alioli:2008tz}
  S.~Alioli, P.~Nason, C.~Oleari and E.~Re,
  JHEP {\bf 0904} (2009) 002
  [arXiv:0812.0578 [hep-ph]].



\bibitem{Hamilton:2009za}
  K.~Hamilton, P.~Richardson and J.~Tully,
  JHEP {\bf 0904} (2009) 116
  [arXiv:0903.4345 [hep-ph]].



\bibitem{Alioli:2009je}
  S.~Alioli, P.~Nason, C.~Oleari and E.~Re,
  JHEP {\bf 0909} (2009) 111
  [Erratum-ibid.\  {\bf 1002} (2010) 011]
  [arXiv:0907.4076 [hep-ph]].





\bibitem{Nason:2009ai}
  P.~Nason and C.~Oleari,
  JHEP {\bf 1002} (2010) 037
  [arXiv:0911.5299 [hep-ph]].


\bibitem{Corcella:2000bw}
  G.~Corcella {\it et al.},
  JHEP {\bf 0101} (2001) 010
  [arXiv:hep-ph/0011363].


\bibitem{Corcella:2002jc}
  G.~Corcella {\it et al.},
  arXiv:hep-ph/0210213.

\bibitem{Sjostrand:2006za}
  T.~Sjostrand, S.~Mrenna and P.~Z.~Skands,
  JHEP {\bf 0605} (2006) 026
  [arXiv:hep-ph/0603175].

\bibitem{Bahr:2008pv}
  M.~Bahr {\it et al.},
  Eur.\ Phys.\ J.\  C {\bf 58}, 639 (2008)
  [arXiv:0803.0883 [hep-ph]].



\bibitem{Frixione:1995ms}
  S.~Frixione, Z.~Kunszt and A.~Signer,
  Nucl.\ Phys.\  B {\bf 467} (1996) 399
  [arXiv:hep-ph/9512328].

\bibitem{Boos:2001cv}
  E.~Boos {\it et al.},
  arXiv:hep-ph/0109068.



\bibitem{Alioli:2010xd}
  S.~Alioli, P.~Nason, C.~Oleari and E.~Re,
  JHEP {\bf 1006} (2010) 043
  [arXiv:1002.2581 [hep-ph]].

\bibitem{PDG}
  http://pdg.lbl.gov


\bibitem{Alwall:2006yp}
  J.~Alwall {\it et al.},
  Comput.\ Phys.\ Commun.\  {\bf 176} (2007) 300
  [arXiv:hep-ph/0609017].

\bibitem{POWHEG_Zjet}
S.~Alioli, P.~Nason, C.~Oleari, and E.~Re, 
In preparation.

\end{thebibliography}
\end{document}